\documentclass{article}

\usepackage{fullpage}

\begin{document}
\title{A Software Architecture Teacher's Dilemmas \footnote{This is the reviewed version of the paper submitted to International Conference on Software Architecture (ICSA 2021)}}
\author{Arvind W Kiwelekar \\
  Department of Computer Engineering\\
  Dr. Babasaheb Ambedkar Technological University \\
Lonere-Raigad India\\
awk@dbatu.ac.in}

\maketitle

\begin{abstract}
An instructor teaching a course on Software Architecture needs to be more reflective to engage students productively in the learning activities.
In this reflective essay, the author identifies a few decisive moments referred to as {\em instructional dilemmas} at which a teacher reflects upon choices and their consequences so that meaningful learning happens.  These situations are referred to as  {\em dilemmas} because they offer two options to instructors.  Some of these dilemmas arise from the inherent nature of Software Architecture as a discipline, while the source of others is the background knowledge of learners. The paper suggests a set of principles and {\em small-teaching methods} to make teaching and learning more effective in such situations.
\end{abstract}

{\bf Keywords:}
Instruction Design and Sequencing,
Instructional Decisions,
Software Architecture Education,
Soft skills for Software Architects.

\section{Introduction}
A course on Software Architecture is increasingly becoming an integral part of the curriculum for the graduate programs in Software Engineering and Computer Science.  
Many academic institutes  are preferring   a dedicated course on Software Architecture over offering it as a  
one of the module in a course on Software Engineering.
This is because of the emphasis of architectural issues in software development processes\cite{angelov2017designing}, and lack of sufficient coverage given to architectural problems in a course on Software Engineering.
 
Such a dedicated course on Software Architecture is usually offered
at the senior undergraduate level
(i.e. third or final year of a four years duration program)
or at the graduate level. 
While offering a course on Software Architecture, it is ensured that students enrolled for  the course
have  completed  prerequisite courses on  Software Engineering, Programming Languages,
Databases, Operating Systems and Distributed Systems.

With this  prerequisite knowledge, the students enrolled in the course
need to understand software design issues such as information storage, transaction management, process concurrency, information presentation, application security and deployment models.
Students are aware of the principles of {\em programming-in-small} \cite{fayad2000thinking,deremer1975programming} such as
data abstraction, information hiding,  code-reuse, and to some extent design reuse through Object-Oriented design patterns.

With the experience of {\em programming-in-small} that too
in only one programming language (e.g., C/Java) students find it difficult
to grasp the high-level systemic issues 
(e.g., Viewpoints, Architectural Styles, Architectural Decisions and their
consequences) dealt in a 
course on Software Architecture. These difficulties stem from two sources. First,
students lack experience of working on a large-sized project. The second source is the abstract nature of   Software Architecture \cite{galster2016makes}.
As a result, students become passive learners, and they start loosing
interest in the course content.

Hence, to engage students more productively during class-room interaction
and as well as through off-class assignments, become a major goal of  teaching. To attain this instructional objective, a teacher needs to make
the right instructional decisions. This paper identifies a few such decisive moments referred to as {\em instructional dilemmas}(Section III) which offers two choices to
a teacher about sequencing of the course content.
The author of the paper  has experienced these situations while teaching courses on Software 
Architecture to undergraduate and graduate students during the last five years and while conducting teacher's training programs.

The  experiences  described  in the paper 
can be useful to the first-time teachers of Software Architecture to design their instructions in a more meaningful way.  Also, the experienced teachers delivering the course can use these experiences to set up controlled experiments for evaluating the effectiveness of their individual choices. Besides, the paper provides a set of guidelines (Section II)
based on modern learning theories which can be useful to deal with similar situations
in either a course on Software Architecture or any other Software Engineering courses. 

\begin{table*}[t]
    \centering
    \begin{tabular}{|p{0.3in}|p{2.5in}|p{0.3in}|p{2.5in}|}
    \hline 
    \multicolumn{2}{|p{2.8in}|}{\centering {\bf Abstractions First}} &\multicolumn{2}{|p{2.8in}|}{\centering {\bf Implementation First}} \\ \hline 
    Unit & Topic & Unit &  Topic \\ \hline      
    1&     {\bf Requirement Analysis}:Architectural Design and Requirements, Allocation & 1& {\bf Architecture Recovery} Documenting Module Dependency View in SEI framework . UML Notations for  Module View \\ \hline 
     2&     {\bf Software Structure and Architecture}: Architectural Structures and Viewpoints, Architectural Styles, design
patterns, architecture design decisions.
  & 2& {\bf Technological Architectures} Service Oriented Architecture, 3 tier Web Application Development (LAMP vas MEAN), Mobile Application Development, Blockchain Technology, Microservices  \\ \hline 
      3&   {\bf Technological Architectures} Service Oriented Architecture, 3 tier Web Application Development (LAMP vas MEAN), Mobile Application Development, Blockchain Technology, Microservices  & 3& {\bf Requirement Analysis}:Architectural Design and Requirements
Allocation \\ \hline 
       4&   {\bf Architecture Evaluation:} Quality Attributes, Quality Analysis and
Evaluation Techniques, Measures & 4& {\bf Software Structure and Architecture}: Architectural Structures and Viewpoints, Architectural Styles, design
patterns, architecture design decisions. \\ \hline 
      5&   {\bf Architecture Recovery and Description}  Documenting Module Dependency View in SEI framework . UML Notations for  Module View & 5&  {\bf Architecture Evaluation:} Quality Attributes, Quality Analysis and
Evaluation Techniques, Measures  \\ \hline 
    \end{tabular}
    \caption{Two Different Course Sequencing Models} \label{2seq}
\end{table*}

 \section{Guiding Principles}
  The guiding principles explained below help us to deal with the instructional dilemmas, to define learning outcomes, and to adopt appropriate teaching methods.
 \begin{enumerate}
  \item {\bf Increase the level of students engagement.}Increased students engagement is typically used as the indicator of measuring learning progress and  effectiveness of course delivery. Designing purposeful learning activities and instruction sequences are the common means through which student's participation can be enhanced \cite{mauricio2018systematic}.  
  \item {\bf Engage students in the knowledge construction process.}  In the conventional view,  teaching is perceived as a process of {\em knowledge dissemination} in which students passively absorb the knowledge. In contrast to this, the {\em constructivist approach} perceives learning as a knowledge construction process in which students develop insight about a topic through the active participation in learning activities. Few educators have highlighted the importance of the constructivist approach while teaching courses on Software Engineering \cite{cain2016reflections}.
  \item {\bf Engage students in Project or Problem based Learning}  Working collaboratively in a team is an essential soft-skill required for engineers in general and  software architects in particular.   Educators suggest designing curriculum and courses around a project or problem-based learning activities to train students for collaborative working \cite{dolog2016assessing,ouh2019applying}.
 \item {\bf Teach concrete things before abstract things.}
 The abstract or vague nature of Software Architecture is the prime reason that makes teaching  Software Architecture difficult \cite{galster2016makes}. Also, education psychologists suggest that teaching concrete things before abstract concepts improve student's comprehension of the subject matter \cite{moreno2011teaching}.
   \item {\bf Adopt of modern technologies.} Modern technologies such as web-based course delivery platforms provide timely and effective media to disseminate course content, engage students in discussions, conduct exit quizzes and seek feed-backs. These technologies can be effectively adopted to address challenges faced by teachers and students.
 \end{enumerate}
The principles mentioned above guide an instructor to devise appropriate interventions to overcome the following instructional dilemmas.

 \section{Instructional Dilemmas}
 This section describes four kinds instructional dilemmas.  These are: (i) Architecture Modelling Dilemma, (ii) Definitional Dilemma,
 (iii) Architecture Design Dilemma, and
 (iv) Implementation Dilemma.
 
 We use a standard template to describe each dilemma. The template includes the following elements.  
 
 \begin{enumerate}
     \item {\bf Context:} This  element describes the background and source of the dilemma .
     \item {\bf Alternatives:} This element describes the two conflicting or competing alternatives.
     \item {\bf Decision:} This element describes the selected alternatives.
     \item {\bf  Guiding Principles:} The rationale behind the selected option is explained. The explanation is based on one or more guiding principles presented in the previous section.
     \item {\bf  Teaching Method:} A teaching method used to deal with the instructional dilemma is described.
     \item {\bf  Learning outcomes:} The expected learning outcomes that will be attained by the application of the teaching method described in this section.
 \end{enumerate}
 The use of a standard template to document the instructional dilemmas is inspired by the framework used to document architectural decisions \cite{tyree2005architecture}.
 
 \subsection{Architecture Modelling Dilemma}
 
 \subsubsection{Context}Software architects usually perform architecture modelling in two different contexts, i.e. forward engineering and reverse engineering.
 During forward engineering,  architects specify a software solution satisfying the given functional requirements and quality attributes. The specified architecture acts as a blueprint for the downstream engineering activities such as low-level design, system implementation, and testing. These kinds of architecture models are referred to as {\em prescriptive} architectures.
During reverse engineering, architects recover high-level descriptions from the implementation to support system maintenance and evolution activities.  These kinds of architecture models are referred to as {\em descriptive} architectures \cite{heldal2016descriptive,eliasson2015architecting}.

\subsubsection{Alternatives}  Whether to begin the course with the descriptive or prescriptive architecture can be the first dilemmatic situation.   The conventional life-cycle model prescribes that the course shall start with the topic of prescriptive architecture.  This sequence of instruction will follow the path in which the first topic would be identifying architecturally significant requirements followed by high-level system design and architecture evaluation.   This instructional sequencing referred to as {\em abstractions first} is shown in the first column of Table \ref{2seq}.   The second alternative is shown in the second column of Table \ref{2seq}. With this sequence,   the course begins with the topic of architecture recovery followed by documenting the extracted architecture with the module view from the SEI framework \cite{clements2003documenting}. The sequencing in the second column is referred to as {\em implementation first} approach.

\subsubsection{Decision} We observe that starting the course with the descriptive architectures is a better choice. 

\subsubsection{Guiding Principles} The following guiding principles justify the choice. (i) Teach concrete concepts before abstract concepts.  (ii) Use technology-based teaching methods to flip the classroom activities.
\subsubsection{Teaching Method}A course typically begins with handing over the code of an open-source software application (e.g., DSpace \cite{smith2003dspace}, HealthWatcher \cite{pinto2007towards}) to students and asking them to document the system with UML notations such as class diagram. Students will struggle to cope with the size of the program when they start documenting such a large-sized application.

The teaching method is based on project-based learning and flipped classroom with the following steps.
\begin{enumerate}
    \item [i] An instructor shares a video explaining Module View from SEI Documentation framework. 
    \item [ii] In the classroom or laboratory session, an instructor helps students to identify dependency relationships among programming elements manually.
    \item [iii]  A group of 3 to 4 students are asked to prepare a report documenting the module view for the assigned system.
\end{enumerate} 
\subsubsection{Learning Outcomes} The students will realize the necessity of higher-level abstractions to gain control over a large-sized application. The students will be able to document a given application in module view.

 \subsection{Definitional Dilemma}
 
 \subsubsection{Context} This dilemma arises from the definition of {\em Software Architecture} as a concept. Two dominant views exist concerning the definition of Software Architecture. The first and classical view suggests that Software Architecture refers to the fundamental structure of a software system manifested through program elements and relationships among them\cite{parnas1972criteria,taibi2019monolithic}. This definition highlights architecting as a {\em decomposition process}. Secondly, one of the most recent approaches defines software architecture as a set of hard decisions made during the initial stages of development that affect quality attributes or externally visible properties. These decisions are usually referred to as {\em architectural decisions} and difficult to change in the later stages of development—for example, decisions concerning the choice of technological platforms or choice of architecture style \cite{tyree2005architecture}.  This definition highlights that software architecture is a {\em decision making} process.
 \subsubsection{Alternatives} These two equally dominant views of Software Architecture provides two alternatives to teach architecture documentation, i.e. whether to teach how to document the decomposition of a system or to teach how to document architectural decisions.  The researchers have developed frameworks to document both,i.e. architectural decisions \cite{ van2012documentation}, as well as, system decomposition \cite{clements2003documenting}. Resolving upon {\em what to document?} becomes a crucial instructional decision once it is decided to emphasize descriptive modelling.
 
 \subsubsection{Decision} We suggest that asking students to document a decomposition of a system is a better choice over architectural decisions. 
 \subsubsection{Guiding Principle} The choice of emphasizing the decomposition view is guided by following observations. (i) Documenting the decomposition of a system is more {\em concrete} activity as compared to documenting architectural decisions. This observation holds, mostly when students are engaged in the architecture recovery process. Because recovering architecture decisions is challenging when information necessary to extract decisions is absent in the implementation  artefacts, on the other hand, information about dependencies among program elements is present and visible in the implementation artefacts, thus simplifying the extraction of high-level views. (ii) Instructors can easily design projects and form teams to support project-based learning when students have to document decomposition of a system.  
  \subsubsection{Teaching Method} A combination of the flipped classroom and project-based learning method can be adapted to emphasize the significance of the structural aspect of software architecture. The teaching method includes following steps.
  \begin{enumerate}
      \item[i]  The instructor shares a video explaining various concerns handled in software design in general which includes concurrency, control, handling of events, data persistence, distribution of components, error and exception handling, interaction, presentation,  and security.
\item[ii] In a lab session, the instructor explains how language-specific mechanisms (e.g., Java) that deal with these concerns
\item[iii] Students are asked to look for the programming elements handling these concerns in the application software assigned to them.
\item[iv] Further, students are asked to group the programming elements handling similar concerns and to prepare a report.
  \end{enumerate}
\subsubsection{Learning Outcomes}
The students will realize that there exist multiple ways to group or decompose a software system. Further, students will realize that there exist few concerns which can not be cleanly grouped called cross-cutting concerns (e.g., Error handling, persistence). The students will be able to identify design concerns in the system implementation and document them with a given template or a UML profile.

 \subsection{Architecture Design Dilemma}
 
 The third kind of dilemma   is concerned with the process of architecture design. The design process   includes two steps. The first stage aims to identify Architecturally Significant Requirements (ASR) \cite{chen2012characterizing} from software requirement specification or a problem statement.  The second step maps architecturally significant requirements to software architecture elements.  
 
 The Pattern-Oriented Software Architecture (POSA)\cite{buschmann2007pattern} and the Attribute-Driven Design (ADD) \cite{bachmann2001introduction} are  two     commonly used  approaches in architecture-centric software development.
 The POSA approach suggests using a ready-made solution to recurrently occurring design problem, which is catalogued as a pattern (e.g., Model-View-Controller, Pipe-and-Filter). The ADD approach is a recursive design method which, when applied, leads to a solution specific to a quality attribute (e.g., Modifiability, Security).
   \subsubsection{Alternatives}The  ADD and POSA are  two approaches available for teaching architecture design. The ADD   as compared to POSA is more systematic and a planned method. The architecture design is a process of solving a {\em wicked problem}; hence,  both approaches lead to multiple and different solutions.  Moreover,   knowing architecture styles and patterns is a prerequisite to apply ADD for solving design problems.
 \subsubsection{Decision} We observe that teaching architecture design with POSA approach followed by a brief explanation of ADD is a better option.
 \subsubsection{Guiding Principle} The students of engineering programs are familiar to solving problems.  As architectural styles and patterns catalogue solutions to recurrent problems,  it will be easier to devise a teaching method around the POSA approach. An instructor can frame a set of questions for a given design problem which can lead to an architecture-style-based solution. Thus students get engaged in the knowledge construction process as recommended by the constructivist theory of learning.
  \subsubsection{Teaching Method} Software architecture design is a specialized skill set of the general-purpose skill called {\em design thinking}. Though all engineering graduates are expected to acquire design thinking, it has been observed that teaching and learning design skill is hard. Various specialized pedagogy centred around {\em project-based learning}  have been devised, and educators have debated their effectiveness  \cite{dym2005engineering}. We suggest adopting the following teaching method aimed at increasing students engagement.
  \begin{enumerate}
      \item[i] The instructor hands over an architecture design problem from a domain to which students are familiar with. For example,  we used problems such as student's academic record management,  room-allotment in student's hostel, travel-grant approval and disbursement. 
      \item[ii] The instructor explains a specific architecture style such as Model-View-Controller (MVC), Pipe-Filter, Client-Server, Service-Oriented Architecture, Micro-Service, Publisher-Subscribers.
      \item[iii] The instructor asks questions that will lead to the identification of Architecturally Significant Requirements.
      \item[iv] The instructor forms multiple teams out of the students attending the class.
      \item[v] The instructor asks  a  team to select an architectural style and to map the ASRs  to architectural elements from the chosen style (e.g., Publisher-Subscriber)
      \item[vi] The instructor displays each groups' mappings on the board so that a solution is visible to everyone.
      \item[vii] Multiple design mappings will be displayed on the board. At which point, the instructor asks the students to select the 'best' mapping and justify their selection.
      \item[viii] At this point, the instructor introduces the concept of architecture evaluation against quality attributes to the students.
  \end{enumerate}
 \subsubsection{Learning Outcomes}During the course delivery an instructor will observe following outcomes:\begin{enumerate}
     \item Increased participation in architecture design activity.
     \item Students will understand the wicked nature of architecture design problem.
     \item Students will be able to apply various architecture styles.
     \item Students will be able to evaluate a given solution against a set of quality attributes.
 \end{enumerate}
\subsection{Implementation Dilemma}
\subsubsection{Context}
The implementation dilemma is not an instructional dilemma  in the sense as described in the previous sections which affect the sequence of instructions.  But, this dilemma  helps to describe  an important topic missed in the course content i.e. {\em architectural decisions}.

A software architect needs to select a particular technology platform to implement the designed architectural solution. For example,  a  client/server web  application  can be realized using either the Linux-Apache-MySQL-PHP (LAMP) stack or the MongoDB-Express.js-Angular.js-Node.js (MEAN) stack. Similarly, the Service-Oriented-Architecture (SOA) and Micro-services are two  technologies to implement  component-oriented distributed applications. These implementation dilemmas are best examples to explain the significance of architectural decisions and documentation  framework for architectural  decisions as defined in \cite{tyree2005architecture}.

 \subsubsection{Alternatives} Technological platforms such as  (i) LAMP vs MEAN (ii) SOA vs Micro-Services (iii) Ethereum vs HyperLedger.
 \subsubsection{Decision} To explain  architectural  elements in various technological platforms at  conceptual level  with an emphasis on quality attributes supported by them. 
 \subsubsection{Guiding Principle} To adopt latest technologies to implement an architectural solution.
  \subsubsection{Teaching Method} The prescribed teaching method is a combination of  classroom lectures and  project based learning. The method includes following instructional steps.
  \begin{enumerate}
      \item[i] The instructor explains the basic building blocks in the following technological platforms (i) LAMP and MEAN (ii) SOA and Micro-Services (iii) Hyperledger and Ethereum
      \item[ii] The instructor asks student to select an appropriate technology to implement the  architectural solution for the problem  used  during architecture design activity. 
      \item[iii] Further, the instructor asks student to document their decision using the framework described in \cite{tyree2005architecture}.
  \end{enumerate}
  
 \subsubsection{Learning Outcomes} The students will develop an insight about architectural decisions and they will be able to document their rationale.

\section{Earlier Work and Discussion}

Teachers and educators from the disciplines other than Software Architecture have identified teaching dilemmas concerning to their domains. For example,  teachers teaching  Mathematics courses experience a conflict between teaching symbolism and logic versus procedural aspects of problem-solving \cite{kamaruddin2012dilemma}.  Further, in Reference \cite{swan2011designing}, the author identifies a set of classroom-teaching dilemmas for Mathematics teachers and asks them to weigh their classroom time devoted to the activity. For example, Mathematics can be learnt either through individual practice or by solving a problem in a team.  

Another example of the teaching dilemma experienced by an English teacher is described in \cite{smagorinsky2011teaching}, which reports the conflict between authoritarian teacher-centric approach versus student-centric approach while teaching grammar and writing.

In this paper, the author identifies a set of conflicting situations experienced while teaching a course on Software Architecture to graduate, undergraduate students and conducting teacher's training program. Unlike earlier approaches described above, the author goes beyond reporting conflicts and tensions among teaching alternatives. In this paper, the author specifies what it is called as {\em small teaching methods} \cite{lang2016small} by James Lang. These small teaching methods are derived from modern learning theories and applicable to day-to-day classroom teaching.

Here, it needs to be noted that these are the dilemmas experienced by an instructor and not  to be confused with conflicting situations encountered by the Software Developer (e.g., particularly implementation dilemma).  These conflicting situations need to be resolved, and their choice affects instruction design and sequence and not the quality of software.

Further, an instructor may experience other kinds of dilemmas such as   breadth versus depth of a concerning topic. Hence, it is not a comprehensive catalogue of teaching dilemmas for a course on Software Architecture.

\section{Conclusion and Future Scope}
Teaching design skills in general and Software Architecture, in particular, has been recognized as a difficult task by many educators.   Industrial professionals have also found that software design as a  skill is difficult to transfer from an expert mentor to a novice trainee joined in the profession because the design and architecting skills heavily make use of experiences accrued over the period. However,  the discipline of Software Architecture has a rich and evolving knowledge base that codify these experiences in the form of  Architectural Styles, Design Patterns,  design methods, and documentation frameworks which when properly utilized can help instructors to design a course instruction aimed to develop software design thinking. 

This paper contributes by identifying some of the challenges faced by instructors. These challenges are referred to as {\em instructional dilemmas}.  Further, the paper contributes by suggesting small-teaching methods specific to a challenge or an instructional dilemma. These small-teaching methods are based on the proven principles in modern learning theories. The author has used these methods to overcome the passivity observed among students and participants attending the teacher's training programs. 

Instructors can use the teaching methods and instructional dilemmas described in the paper to design and plan instruction sequence. Also, the effectiveness of these methods needs to be evaluated in terms of attainment of course objectives.
 \bibliographystyle{IEEEtran}
\bibliography{paper}
\end{document}